\documentclass[osajnl,preprint,showpacs]{revtex4}
\DeclareRobustCommand{\baselinestretch{2}}

\usepackage{graphicx}

\begin{document}

\title{Numerical study of light correlations in a random medium close to Anderson localization threshold}

\author{Shih-Hui Chang}
\affiliation{Department of Electrical and Computer Engineering, Northwestern University, Evanston, IL 60208}

\author{Allen Taflove}
\affiliation{Department of Electrical and Computer Engineering, Northwestern University, Evanston, IL 60208}

\author{Alexey Yamilov}
\affiliation{Department of Physics and Astronomy, Northwestern University, Evanston, IL 60208}

 \author{Aleksander Burin}
\affiliation{Department of Physics and Astronomy, Northwestern University, Evanston, IL 60208}

\author{Hui Cao}
\affiliation{Department of Physics and Astronomy, Northwestern University, Evanston, IL 60208}

\email{a-yamilov@northwestern.edu}

\date{\today}

\begin{abstract}
We applied finite difference time domain (FDTD) algorithm to the study of field and intensity correlations in random media. Close to the onset of Anderson localization, we observe deviations of the correlation functions, in both shape and magnitude, from those predicted by the diffusion theory. Physical implications of the observed phenomena are discussed. 
\end{abstract}
%\ocis{030.1670,290.4210,290.1990}
\maketitle

%030.1670  Coherent optical effects
%290.4210  Multiple scattering
%290.1990  Diffusion

Light propagating in a random medium undergoes multiple scattering. As a result, it forms a complex interference (speckle) pattern when it emerges from the medium, which can be described statistically\cite{patrick_book,shapiro_ce}. There exist longer range correlations, that can be detected as correlations between two distant speckles\cite{feng_cf_prl,shapiro_cf_prl}. These longer range correlations in transmitted intensities received much attention in context of mesoscopic electronic systems\cite{beenakker}, where they result in universal conductance fluctuations. For electronic systems only limited information (fluctuation of conductance) can usually be deduced from experiments. For light, however, momentum- (angular) and spatial-resolved measurements are possible. The latter yield much more detailed information about the system. The connection between electrons and photons is established by identifying the Landauer conductance $g$ with the total transmission coefficient (summed over all incoming and outgoing channels)\cite{beenakker}. The majority of experimental and theoretical efforts (see Refs. \cite{patrick_book,ping_sheng,feng_phys_rep,rossum_rmp} and references therein) to date are concentrated on systems in the regime of diffusive transport, $g\gg1$. The intensity correlation function (ICF) is defined as\cite{feng_phys_rep,rossum_rmp}
\begin{equation}
C(\Delta r, \Delta \nu)=\frac{\langle I({\bf r}+\Delta{\bf r},\nu+\Delta\nu)I({\bf r},\nu)\rangle}{\langle I({\bf r}+\Delta{\bf r},\nu+\Delta\nu)\rangle \langle I({\bf r},\nu)\rangle}-1,
\label{ci}
\end{equation}
where ${\bf r}$ and $\nu$ are spatial coordinate and frequency respectively. When length $L$ of the random medium is greater that its width $W$, $C$ is invariant with respect to ${\bf r}$, and isotropic\cite{genack_symmetry} for $\Delta {\bf r}$, as long as one avoids the evanescent zone on the output surface\cite{apostol_dogariu}. Theoretically, based on pairing of incoming and outgoing channels, three %\cite{shapiro_c0} 
contributions to ICF have been identified\cite{feng_cf_prl,shapiro_cf_prl,feng_phys_rep,rossum_rmp}: short-range $C_1$, long-range $C_2$ and infinite-range $C_3$. Deep into diffusion regime $g\gg 1$ in a wave-guide geometry\cite{feng_phys_rep,rossum_rmp}  $C_1\simeq |C_E(\Delta r, \Delta \nu)|^2\sim 1$, $C_2\sim 1/g$ and $C_3\sim 1/g^2$, making the values of $C_2$ and $C_3$ small. Here, 
\begin{equation}
C_E(\Delta r, \Delta \nu)=\frac{\langle E({\bf r}+\Delta{\bf r},\nu+\Delta\nu)E^*({\bf r},\nu)\rangle}{\langle I({\bf r}+\Delta{\bf r},\nu+\Delta\nu)\rangle^{1/2} \langle I({\bf r},\nu)\rangle^{1/2}}.
\label{ce}
\end{equation}
is the field correlation function (FCF). Essentially, the nonpertubative nature of the crossover from diffusion to localization limits the amount of information that can be obtained theoretically\cite{mirlin}. The purpose of this letter is to introduce a {\it numerical} method for studying correlation functions (CFs) based on FDTD\cite{taflove} algorithm. This method allows us to solve Maxwell's equations for electromagnetic field at every spatial grid at every time step. It makes no assumption about the scattering strength, accounts for all interference phenomena, and makes it possible to obtain CFs in both diffusion and localization regimes. Here we apply our method to systems with values of $g$ from $4.5$ to $1$.

We consider a 2D system as shown in inset of Fig. 1: parallel plate metallic waveguide filled with circular dielectric scatterers of refractive index $n=2$ and diameter $d=1.4\ cm$. We choose  our parameters close to the microwave experiments of Ref. \cite{bing_hu}. The scatterers were randomly positioned (without overlapping) with a fixed filling fraction $f=0.3$. Metallic walls of the waveguide insured 99.9\% reflectivity, and uniaxial perfectly matched layer\cite{taflove} boundary conditions were applied outside the waveguide. A TM polarized broadband pulse with center wavelength $\lambda=2\ cm$ (close to one of Mie resonances of the scatterers) was launched via a point source placed one wavelength away from the left surface of the random medium inside the waveguide. A temporal discrete Fourier transform (DFT) was applied to electric fields observed at a series of vertical planes separated by $\lambda/2$ at the output end of the waveguide. By virtue of DFT, we obtained the continuous wave response of the system for a number of wavelengths $\lambda_i$ within a 4\% range of $\lambda$. This range was significantly narrower than (i) the width of the broadband pulse, and (ii) the range in which physical parameters of the system start to deviate from those at $\lambda$. The distance between two consecutive $\lambda_i$ was chosen to be equal to the average mode spacing. Spatial field and intensity CFs are obtained by setting $\Delta\nu=0$ in Eqs. \ref{ci},\ref{ce}. We averaged over 48 configurations and 60 $\lambda_i$, similar to the microwave experiments of Ref. \cite{bing_hu}. In addition, due to the invariance of correlations in the empty part of the waveguide, we averaged over 10 observation planes. Up to 100 spatial points equally spaced near the center of each plane were sampled depending on $\Delta r$. For spectral CFs, Eqs. (\ref{ci},\ref{ce}) with $\Delta r=0$,  the averaging procedure is similar except that the number of $\lambda_i$ sampled depends on the value of $\Delta \nu$. 

Waveguide geometry makes our system quasi-1D, where localization length $\xi \propto Nl$. $N$ is the number of waveguide channels, and $l$ is the transport mean free path. With an increase of $L$, the system crosses over from diffusion to localization while $l$ is kept constant. According to the diffusion theory the dimensionless conductance of a quasi-1D system $g=(\pi/2) n_{eff}^{(e)}Nl/L^\prime$, where $L^\prime=L+2z_b$ accounts for the boundary effect\cite{lagendijk}, the extrapolation length $z_b$ is usually of order $l$, $n_{eff}^{(e)}=c/v_E$, $v_E$ is the energy transport velocity \cite{bart_ve}. In a passive system $g$ is equal to the Thouless number which is defined as the ratio of the diffusion mode linewidth to average mode spacing\cite{ping_sheng,andrey_nature}. In our system,  $l=1.8\ cm$, $n_{eff}^{(e)}=1.77$(see below). The width of the waveguide, $W=20\ cm$ ($N=2W/\lambda=20$) and lengths $L=20,40$ and $80\ cm$ yielded the values of $g=4.4,2.2$ and $1.1$ respectively, and allowed us to study the correlation functions near the onset of localization. 

$z_b$ can be determined by fitting the real part of spatial FCF with the formula
\begin{equation}
C_E(\Delta r)=\frac{\pi (z_b/l) J_0(k\Delta r) +2\sin (k\Delta r)/k\Delta r}{\pi z_b/l+2},
\label{ce_dr_theory}
\end{equation}
where $k=2\pi/\lambda$, and $J_0$ is Bessel function of zero order. The imaginary part of $C_E(\Delta r)$ should vanish due to isotropy\cite{genack_symmetry}, which is confirmed by our calculation in which its value was less than $10^{-3}$. We derived the above expression in 2D following the 3D derivation of Ref. \cite{eliyahu_surf}. Eq. (\ref{ce_dr_theory}) gives an excellent fit for all systems studied with $z_b/l=0.8$ in Fig. 1. Absence of the deviation of $C_E(\Delta r)$ from the expression derived in diffusion regime becomes apparent once one recognizes that $C_E(\Delta r)$ contains only information about short range correlations. It reflects the correlations on length scale of $l$, which is much smaller than $\xi$. $n_{eff}^{(e)}$ in the expression for $g$ can be calculated  $n_{eff}^{(e)}=(1\times W_{air}+n\times W_{scat})/(W_{air}+W_{scat})$, where $W_{air}$ and $W_{scat}$ are the energy stored in air and scatterers respectively. $W_{air}$ and $W_{scat}$ were determined numerically to give $n_{eff}^{(e)}\simeq 1.77$. The physical parameter yet to be obtained is the transport mean free path $l$. This can be done by fitting the spectral CF to the complex function\cite{garcia_genack,van_rossum} 
\begin{equation}
C_E(\Delta \nu)=qL^{\prime}/\sinh qL^{\prime},
\label{ce_dnu_theory}
\end{equation}
with $q=\sqrt{\pi \Delta\nu/D}(1-i)$. $l$ enters Eq. (\ref{ce_dr_theory}) through $L^{\prime}$ and the diffusion coefficient $D=v_El/2$. Fig. 2 shows the fitting of the real and imaginary parts of $C_E(\Delta \nu)$, from which we find $l=1.8\ cm$.  The half width at half maximum (HWHM) of $|C_E(\Delta \nu)|^2$ should coincide with $1.46$ times the diffusion mode linewidth $\delta \nu=D/L^{\prime\ 2}$, at least in the diffusion regime. The inset in Fig. 2 shows that the obtained $l$ and $\delta \nu$ are consistent. A slight deviation in the system of $g=4.4$ is attributed to its short length.

The spatial dependence of long-range contribution to ICF was derived in diffusion up to $1/g^2$ order\cite{bing_hu}
\begin{equation}
C(\Delta r)-\left|C_E(\Delta r)\right|^2=\left(\frac{4}{3g}+\frac{8}{15g^2}\right)\frac{1+\left|C_E(\Delta r)\right|^2}{2}.
\label{c_long_dr_theory}
\end{equation}
Fig. 3 shows both magnitude and normalized profile of Eq. (\ref{c_long_dr_theory}). The numerically calculated $C(\Delta r)-\left|C_E(\Delta r)\right|^2$, shown in Fig. 3, reveals that its spatial profile is independent of $g$. Specifically, the ratio of its value at $\Delta r=0$ to that at $\Delta r \rightarrow \infty$ remains equal to two. Stronger oscillations in the numerical data are likely due to the finite width of the waveguide. However, the magnitude of the long-range contribution is significantly enhanced due to localization effects, far beyond the diffusion prediction up to order $1/g^2$.

In Fig. 4, as $g$ decreases, the long range contribution to spectral ICF, $C(\Delta \nu)-\left|C_E(\Delta \nu)\right|^2$, is broadened when $\Delta \nu$ is normalized to $\delta \nu$. We ascribe this effect to strong fluctuations close to the localization threshold. Namely, a few (more conducting) channels with larger than average linewidth dominate ICF, leading to its spectral  broadening.

In conclusion, using the FDTD algorithm, we studied field and intensity correlation functions close to the onset of localization. In this regime neither experiments nor analytical theories have given such detailed information about correlation of intensities transmitted through a random medium.

AY and HC are grateful to Azriel Genack and Boris Shapiro for their comments on the manuscript and fruitful discussions. This work is supported by the National Science Foundation under Grant No DMR-0093949. HC acknowledges the support from the David and Lucile Packard Foundation.

\newpage
{\bf Figure captions}
\begin{description}

\item[Figure 1.] Real part of spatial FCF. Squares, circles and triangles correspond to the systems with $g=4.4,\ 2.2,$ and $1.1$ respectively. Solid curve represents the fit of Eq. (\ref{ce_dr_theory}) with $z_b/l=0.8$. The inset shows the geometry of our system.

\item[Figure 2.] Real (empty symbols) and imaginary (full symbols) parts of frequency FCF. Solid and dashed curves represent real and imaginary parts of $C_E$ given by Eq. (\ref{ce_dnu_theory}) with $l=1.8\ cm$. The inset compares $\delta \nu$ found from HWHM of $|C_E(\Delta \nu)|^2$ to $D/L^{\prime 2}$. Both quantities are normalized to average mode spacing. Symbol notations are the same as in Fig. 1.
  
\item[Figure 3.] The magnitude of long-range contribution to ICF versus dimensionless conductance $g$. Symbol notations are the same as in Fig. 1. Solid line represents diffusion expansion formula Eq. (\ref{c_long_dr_theory}) at $\Delta r=0$. The inset shows long-range contribution to ICF normalized to one at $\Delta r=0$. Solid, dashed and dot-dashed curves correspond to samples with $g=4.4,\ 2.2,$ and $1.1$ respectively. Thick solid curve plots Eq. (\ref{c_long_dr_theory}).

\item[Figure 4.] Frequency dependence of long-range contribution to ICF normalized to value at $\Delta \nu=0$ . Symbol notations are the same as in Fig. 1.
\end{description}

\newpage
\begin{figure}
\centerline{\rotatebox{-90}{\scalebox{0.7}{\includegraphics{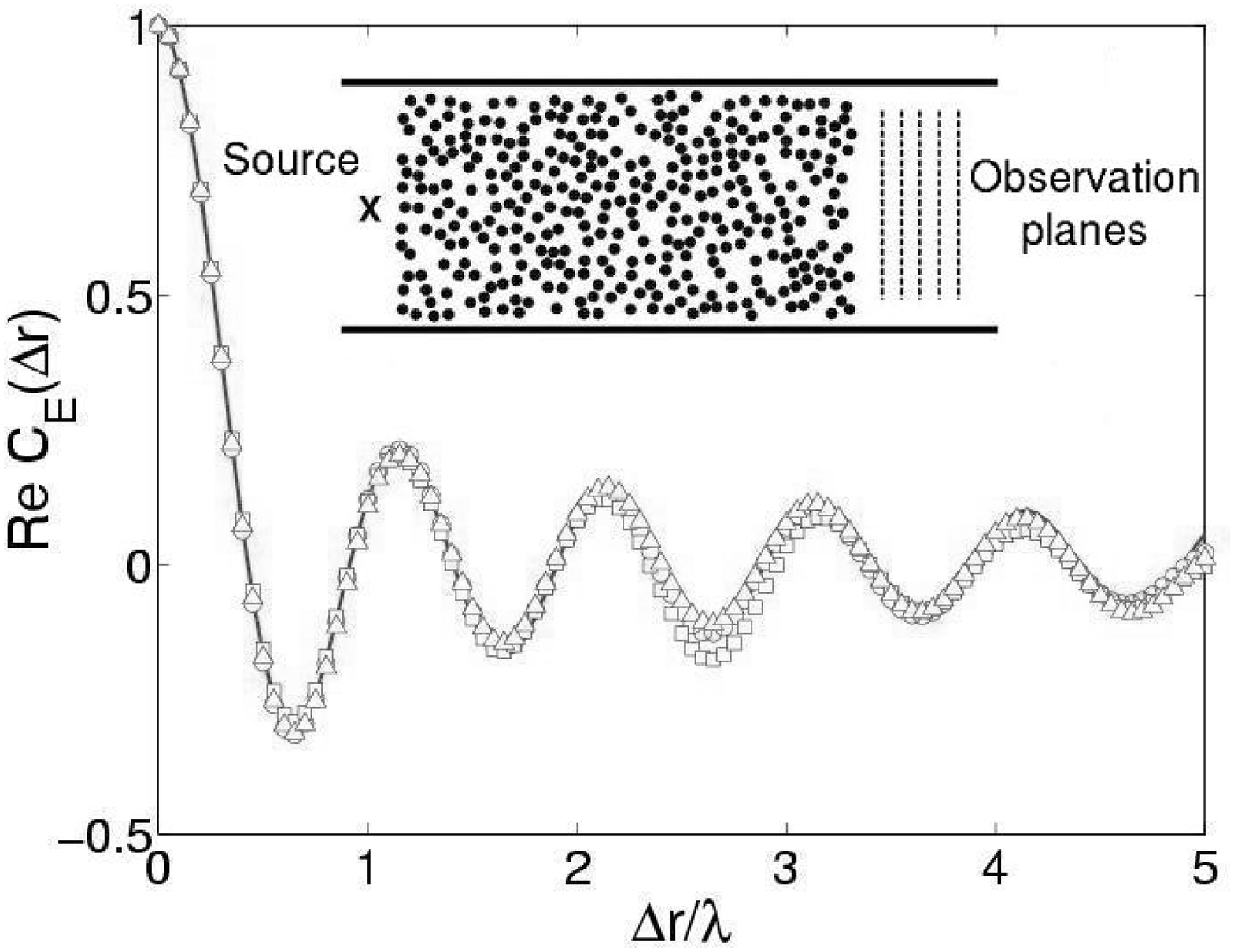}}}}
%\label{ce_dr}
\end{figure}
{\LARGE {\bf Figure 1}}

\newpage
\begin{figure}
\centerline{\rotatebox{-90}{\scalebox{0.7}{\includegraphics{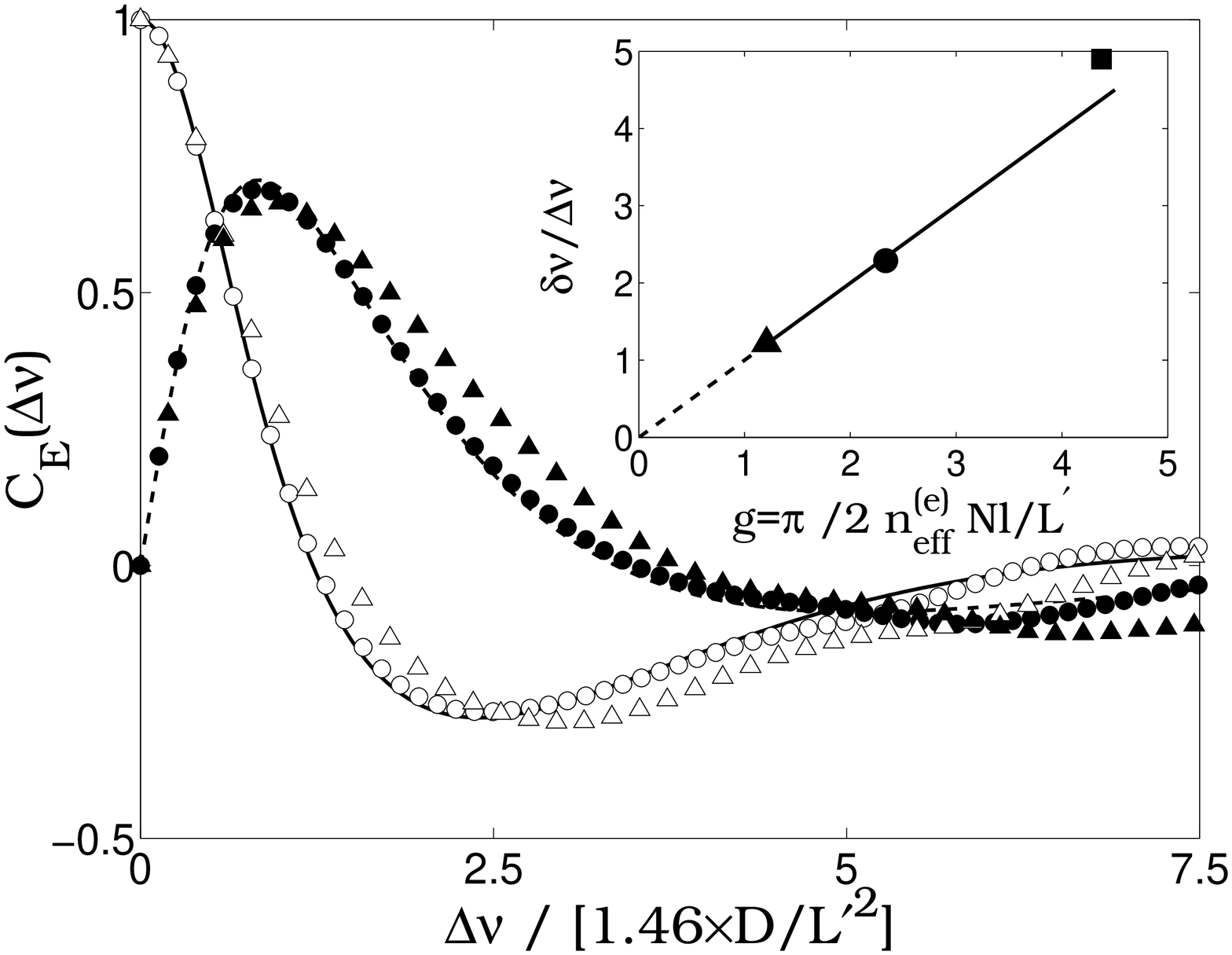}}}}
%\label{ce_dnu}
\end{figure}
{\LARGE {\bf Figure 2}}

\newpage
\begin{figure}
\centerline{\rotatebox{-90}{\scalebox{0.7}{\includegraphics{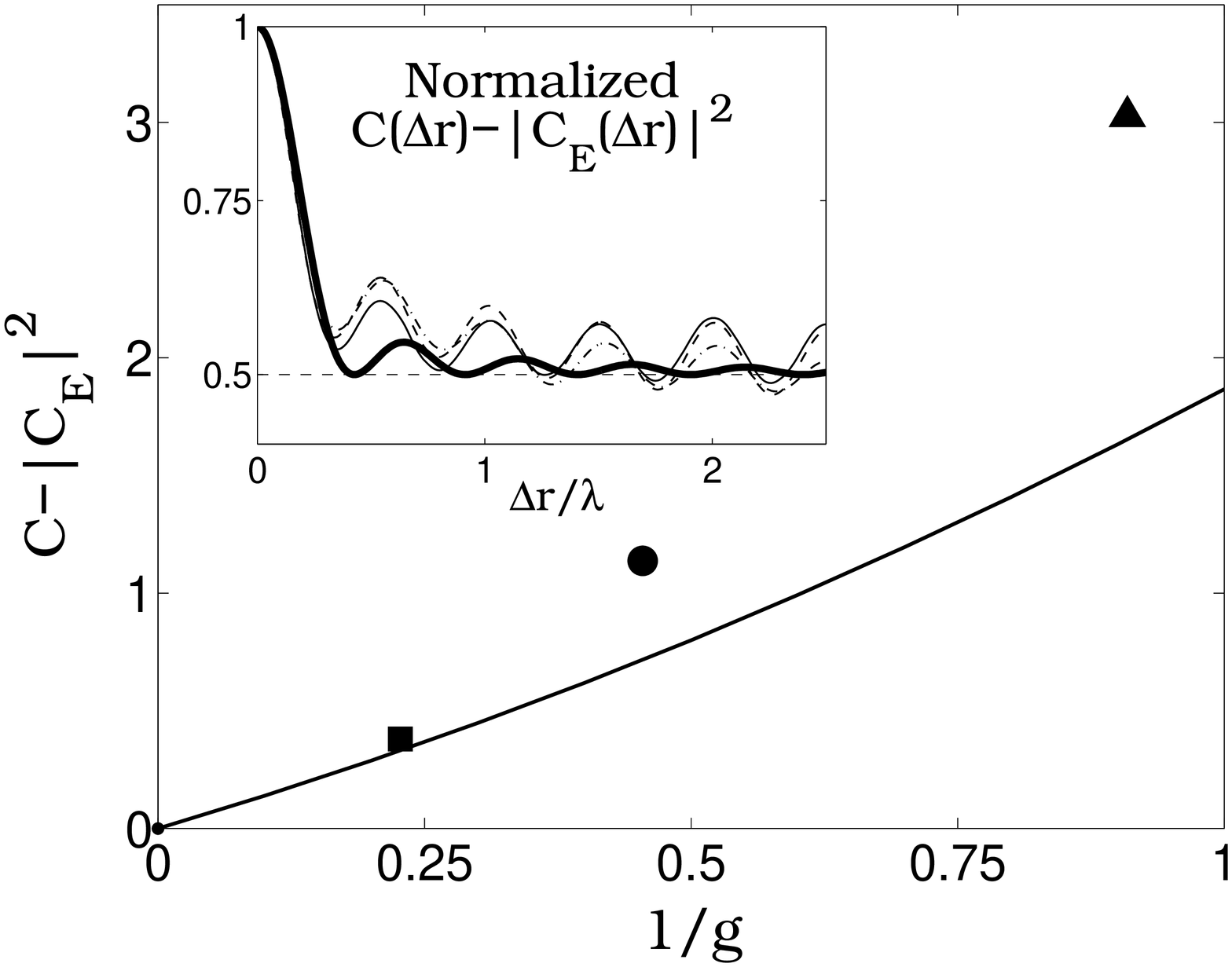}}}}
%\label{c_long_vs_dr}
\end{figure}
{\LARGE {\bf Figure 3}}

\newpage
\begin{figure}
\centerline{\rotatebox{-90}{\scalebox{0.7}{\includegraphics{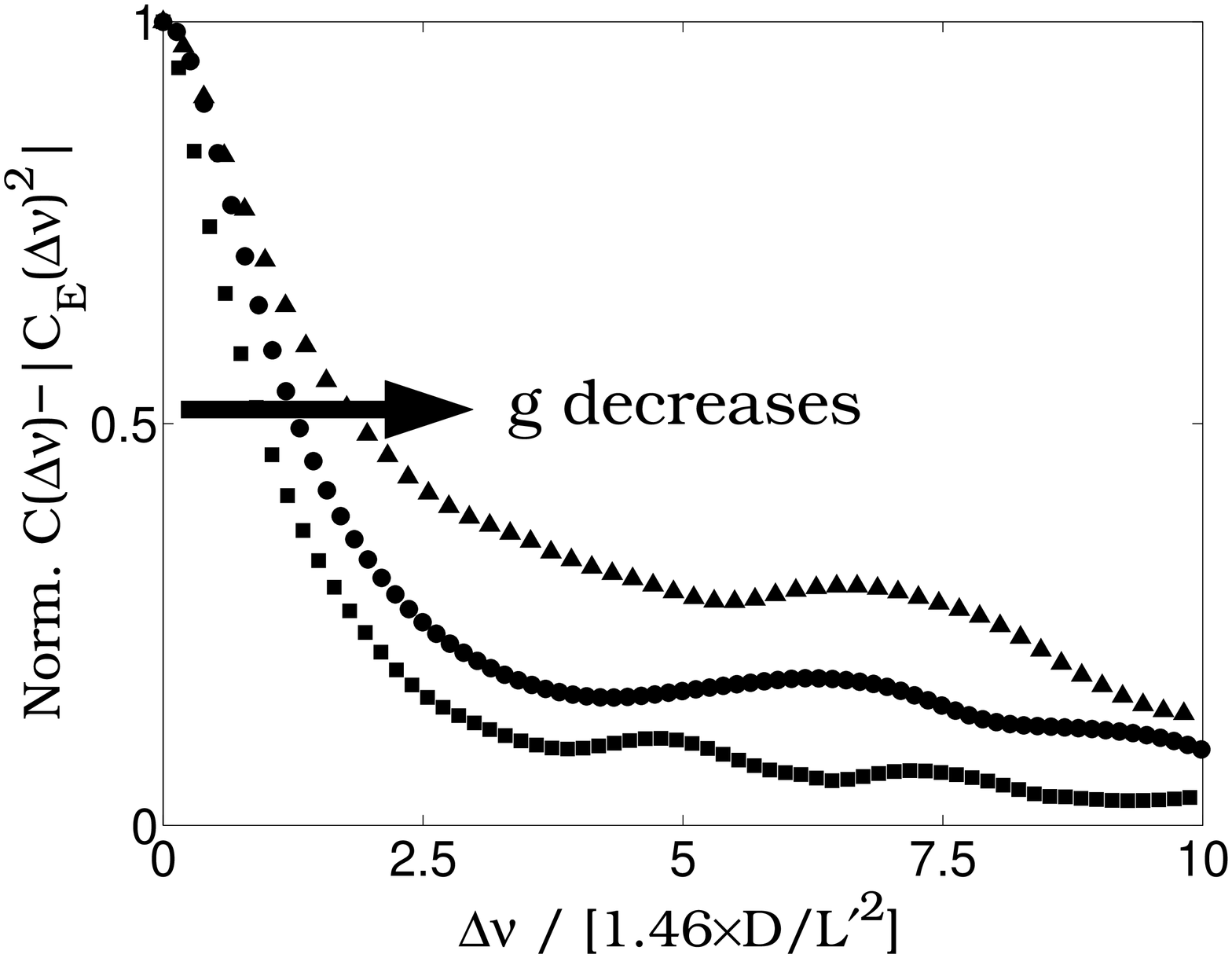}}}}
%\label{c_long_vs_dnu}
\end{figure}
{\LARGE {\bf Figure 4}}
\end{document}